\documentclass[twocolumn]{aastex63}
\received{}
\revised{}
\accepted{}
\submitjournal{ApJ}

\shorttitle{Integrated photometry of multiple stellar populations in Globular Clusters}
\shortauthors{Jang et al.\,}

\begin{document}

\title{Integrated photometry of multiple stellar populations in Globular Clusters.}

\correspondingauthor{Sohee Jang}
\email{sohee.jang@unipd.it}

\author[0000-0002-1562-7557]{S.\,Jang}
\affiliation{Dipartimento di Fisica e Astronomia ``Galileo Galilei'', Universit\`{a} di Padova, Vicolo dell'Osservatorio 3, I-35122, Padua, Italy}

\author[0000-0001-7506-930X]{A.\,P.\,Milone}
\affiliation{Dipartimento di Fisica e Astronomia ``Galileo Galilei'', Universit\`{a} di Padova, Vicolo dell'Osservatorio 3, I-35122, Padua, Italy}
\affiliation{Istituto Nazionale di Astrofisica - Osservatorio Astronomico di Padova, Vicolo dell'Osservatorio 5, IT-35122, Padua, Italy}

\author[0000-0003-1713-0082]{E.\,P.\,Lagioia}
\affiliation{Dipartimento di Fisica e Astronomia ``Galileo Galilei'', Universit\`{a} di Padova, Vicolo dell'Osservatorio 3, I-35122, Padua, Italy}

\author[0000-0002-1128-098X]{M.\,Tailo}
\affiliation{Dipartimento di Fisica e Astronomia ``Galileo Galilei'', Universit\`{a} di Padova, Vicolo dell'Osservatorio 3, I-35122, Padua, Italy}

\author[0000-0003-1757-6666]{M.\,Carlos}
\affiliation{Dipartimento di Fisica e Astronomia ``Galileo Galilei'', Universit\`{a} di Padova, Vicolo dell'Osservatorio 3, I-35122, Padua, Italy}

\author[0000-0001-8415-8531]{E.\,Dondoglio} 
\affiliation{Dipartimento di Fisica e Astronomia ``Galileo Galilei'', Universit\`{a} di Padova, Vicolo dell'Osservatorio 3, I-35122, Padua, Italy}

\author[0000-0003-2373-0404]{M.\,Martorano}
\affiliation{Dipartimento di Fisica e Astronomia ``Galileo Galilei'', Universit\`{a} di Padova, Vicolo dell'Osservatorio 3, I-35122, Padua, Italy}

\author{A.\,Mohandasan}
\affiliation{Dipartimento di Fisica e Astronomia ``Galileo Galilei'', Universit\`{a} di Padova, Vicolo dell'Osservatorio 3, I-35122, Padua, Italy}

\author[0000-0002-1276-5487]{A.\,F.\.Marino}
\affiliation{Istituto Nazionale di Astrofisica - Osservatorio Astronomico di Padova, Vicolo dell'Osservatorio 5, IT-35122, Padua, Italy}
\affiliation{Istituto Nazionale di Astrofisica - Osservatorio Astrofisico di Arcetri, Largo Enrico Fermi, 5, Firenze, IT-50125}

\author[0000-0002-7690-7683]{G.\,Cordoni}
\affiliation{Dipartimento di Fisica e Astronomia ``Galileo Galilei'', Universit\`{a} di Padova, Vicolo dell'Osservatorio 3, I-35122, Padua, Italy}

\author[0000-0002-2210-1238]{Y.-W.\,Lee}
\affiliation{Center for Galaxy Evolution Research and Department of Astronomy,\\Yonsei University, Seoul 03722, Korea}

\begin{abstract}
Evidence that the multiple populations (MPs) are common properties of globular clusters (GCs) is accumulated over the past decades from clusters in the Milky Way and in its satellites. 
This finding has revived GC research, and suggested that their formation at high redshift must have been a much-more complex phenomenon than imagined before.

However, most information on MPs is limited to nearby GCs. The main limitation is that most studies on MPs rely on resolved stars, facing a major challenge to investigate the MP phenomenon in distant galaxies. 
Here we search for integrated colors of old GCs that are sensitive to the multiple-population phenomenon. To do this, we exploit integrated magnitudes of simulated GCs with MPs, and multi-band {\it Hubble Space Telescope} photometry of 56 Galactic GCs, where MPs are widely studied, and characterized as part of the UV Legacy Survey of Galactic GCs.

We find that both integrated $C_{\rm F275W,F336W,F438W}$ and $m_{\rm F275W}-m_{\rm F814W}$ colors strongly correlate with the iron abundance of the host GC. 
In second order, the pseudo two-color diagram built with these integrated colors is sensitive to the MP phenomenon.
 In particular, once removed the dependence from cluster metallicity, the 
  color residuals depend on the maximum internal helium variation within GCs and on the fraction of second-generation stars.  

This diagram, which we define here for Galactic GCs, has the potential of detecting and characterizing MPs from integrated photometry of old GCs, thus providing the possibility to extend their investigation outside the Local Group.

\end{abstract}


\section{Introduction}\label{sec:intro}
Once considered prototypes of simple stellar populations, Milky Way (MW) Globular Clusters (GCs) are typically composed of multiple populations of stars: a first stellar generation (1G) with similar content of light elements as halo field stars with similar metallicity and a second generation (2G) of stars enhanced in He, N and Na and depleted in C and O  \citep[e.g.][]{carretta2009a, marino2019a}.
 The origin of 2G stars as well as GC formation are among of the main issues of stellar astrophysics. Some scenarios predict that the proto GCs have experienced multiple star-formation bursts. These primordial stellar systems were much more-massive than present-day GCs and have lost the majority of their 1G, thus providing a significant contribution to the assembly of the Galaxy \citep[e.g.][]{dantona2016a, decressin2007a, renzini2015a, denissenkov2014a}. 
 In contrast, other scenarios suggest that all GC stars are nearly coeval and that the bizarre chemical composition of the 2G rises from some exotic phenomena as accreation of polluted material during the pre-main sequence (MS) phase \citep[e.g.][]{demink2009a, bastian2013a, gieles2018a}. 

 First- and second-generation stars define distinct sequences in color-magnitude diagrams (CMDs) and pseudo CMDs built with appropriate magnitudes. In particular, the color difference between 1G and 2G stars with similar luminosity is indicative of their chemical compositions. 
  Hence, photometry of resolved stars has been widely used as a tool to identify and characterize multiple stellar populations in GCs. 

Stellar populations have been recently studied in a large number of GCs by using multi-band {\it HST} photometry. In particular, the $m_{\rm F275W}-m_{\rm F814W}$ stellar colors and the  $C_{\rm F275W,F336W,F438W}$ pseudo-colors are powerful tools to identify and characterize multiple populations in GCs as they are very sensitive to the stellar abundances of nitrogen and helium, respectively \citep[e.g.][]{milone2017a,lagioia2018a}. Multiple populations have been also detected in the GCs of  Milky Way satellites like the Magellanic Clouds and Fornax and in M\,31 GCs. \citep[e.g.][]{martocchia2019a, milone2020a, larsen2014a, nardiello2018a}.

The investigation of MP phenomenon beyond the Local Group is more challenging.
The main limitation is that most studies on multiple populations rely on photometry of resolved stars and it is not possible to resolve GC stars of distant galaxies with present-day facilities.  

Pioneering attempts to search for multiple populations from integrated photometry come from observations of GCs in the Virgo elliptical galaxy, M\,87.
Based on optical, near UV and far UV colors, \citet{sohn2006} and \citet{kaviraj2007} provided possible evidence for GCs with super-He-rich stellar populations.
 New insights are provided by \citet{bellini2015}, based on F275W, F606W and F814W integrated {\it HST} photometry. They detected a difference between the UV-to-optical flux ratio of the innermost and outer regions of some M\,87 GCs. This result is consistent with multi-population GCs where the helium-rich 2G stars are more centrally concentrated than the 1G. 
Although these works represent the state of the art for studies on distant galaxies, possible evidence of multiple populations is provided for a small sample of GCs with extreme properties alone.

Furthermore, abundance analysis from integrated-light spectra have been carried out for extragalactic GCs. Evidence for star-to-star abundance variation in light elements such as Mg, Na, and Al have been inferred within  old and intermediate-age GCs in M\,31, M\,33, in the Large Magellanic Cloud, Fornax dwarf spheroidal, NGC\,147, NGC\,6822, NGC\,1316, and in the WLM dwarf galaxy 
 \citep[][]{bastian2019,colucci2009,colucci2012,colucci2014,colucci2017,lardo2017,larsen2012,larsen2014b,larsen2018,sakari2015,schiavon2013}.

In this work, we search for a new photometric diagram, based on integrated photometry, that is sensitive to multiple populations in GCs.
 To do this, we exploit 
  56 Galactic GCs where multiple populations have been widely studied from resolved stars.  Integrated photometry is  derived  from images collected through the F275W, F336W, F438W 
   and F814W filters of the {\it Hubble Space Telescope}, which are the photometric bands mostly used to identify 1G and 2G stars. 
 Moreover, we analyze multi-band integrated photometry of simulated stellar populations with chemical composition representative of 1G and 2G stars.
The resulting integrated magnitudes of simulated and observed clusters are used to investigate the potential of integrated photometry to identify and characterize multiple stellar populations.

The paper is organized as following.
In Section \ref{sec:data} we describe the observational data and the simulated photometry. 
In Section \ref{sec:3} we examine the integrated photometry of the distinct stellar populations in those GC, where 1G and 2G can be separated along the CMD. In Section \ref{sec:4} we investigate the integrated photometry of 56 galactic GCs and relations with cluster metallicity.
Section \ref{sec:5} searches for a link between integrated colors and the main parameters related to multiple populations in MW GCs. Finally, in Section \ref{sec:6} we summarize the main findings and discuss how the results of this paper would serve as a stepping stone to expand the studies on the multiple stellar populations in GCs onto the distant unresolved extragalactic GCs.  

\begin{figure*} 
\begin{center} 
  \includegraphics[height=9.2cm,trim={0.5cm 0.1cm 9cm 2.8cm},clip]{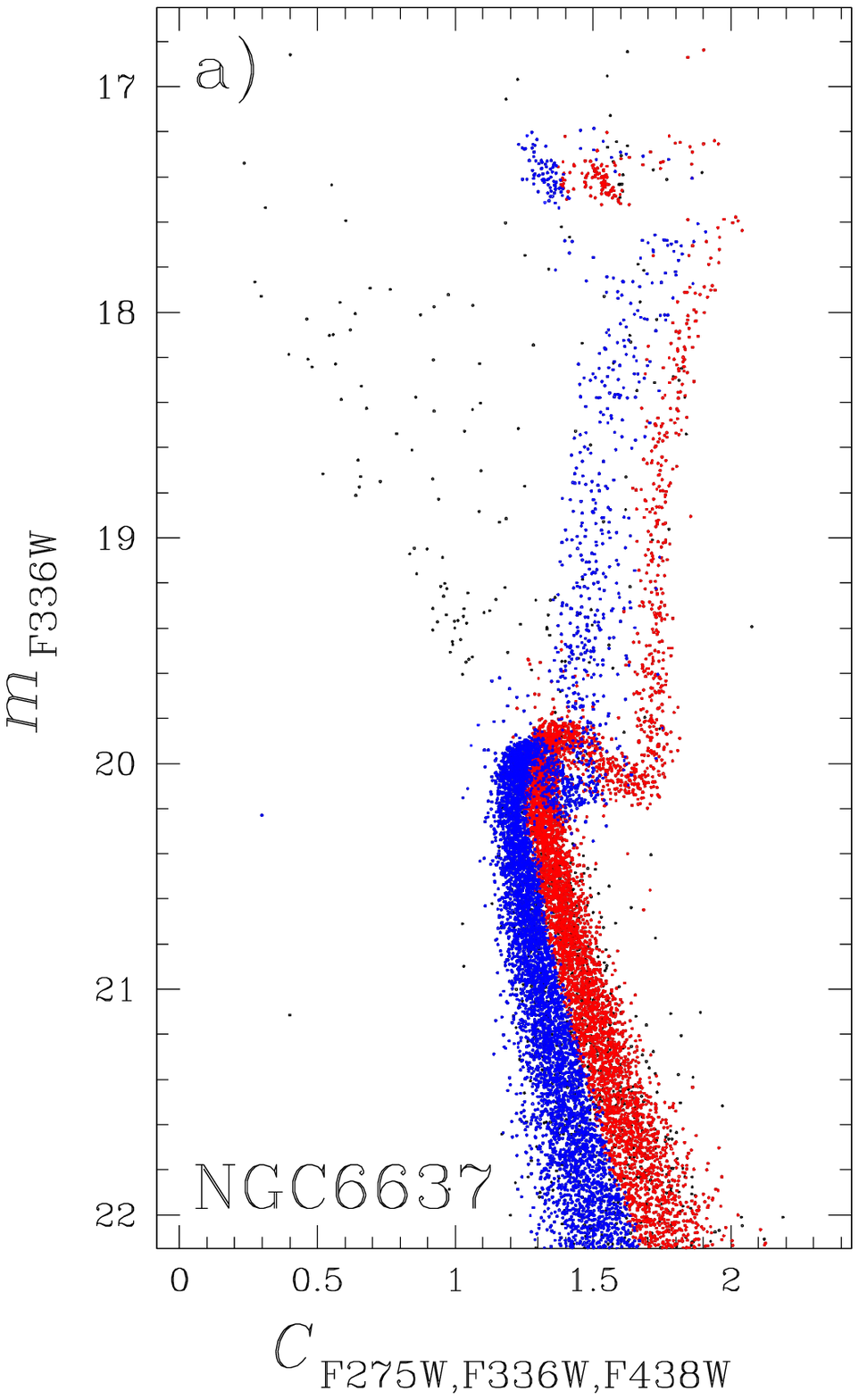}
    \includegraphics[height=9.9cm,trim={0cm 0.5cm 5cm 4cm},clip]{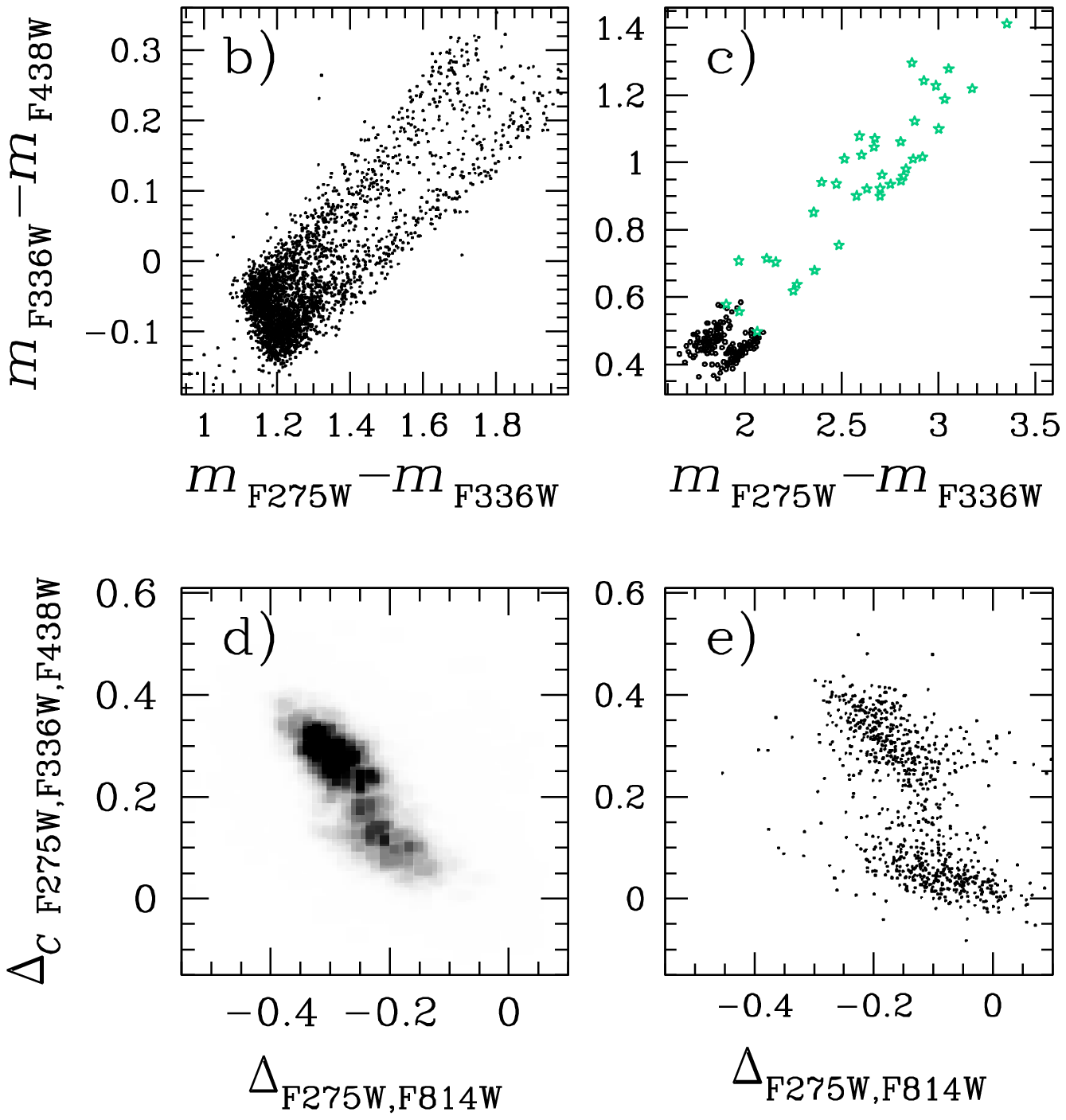}
\caption{\textit{Left panel.} $m_{\rm F336W}$ vs.\,$C_{\rm F275W,F336W,F438W}$ diagram of NGC\,6637. Red and blue points correspond to 1G and 2G stars, respectively. Right panels show the four diagrams used to select SGB stars (panel b), HB and AGB stars represented with black dotted and green starred symbols, respectively (panel c),  MS (panel c) and  RGB stars (panel 6). See text for details.}
 \label{fig:NGC6637}
 \end{center} 
\end{figure*} 

\section{Data and data analysis}\label{sec:data}
To investigate the integrated colors of multiple stellar populations in GCs, we used both {\it HST} observations of 56 Galactic GCs and simulated photometry. Details on these observational and theoretical datasets are provided in the next sections.

\subsection{Observational data}
To derive the integrated fluxes of GC stars we exploited the photometric and astrometric catalogs by \citet{nardiello2018a}, which are based on data collected as part of the Advanced Camera for Survey (ACS) survey of Galactic GCs \citep[GO-10775, PI A.\,Sarajedini,][]{sarajedini2007a} and the {\it HST} Ultraviolet Legacy Survey of Galactic GCs \citep[GO-13227, PI G.\,Piotto,][]{piotto2015a}. The catalogs by Nardiello and collaborators provide photometry of stars in the central $\sim 2.7 \times 2.7$ arcmin field of view in the F275W, F336W and F438W bands of the Ultraviolet and Visual Channel of the Wide-Field Camera 3 (UVIS/WFC3) and the F606W and F814W filters of the Wide-Field Channel of ACS (WFC/ACS). Only stars that according to the proper motions are cluster members (membership probability higher than 0.95) have been used to derive integrated magnitudes. Photometry has been corrected for differential reddening as in \citep{milone2012a}.

\subsection{Simulated photometry}
To qualitatively investigate the integrated colors of stellar populations 
  we have constructed a grid of synthetic stellar evolution models together with HB stars, following the techniques developed by \citet{lee1990,lee1994}, and updated by \citet{jang2014a} and \citet{jang2015} to account for helium-rich multiple populations. Our simulations are based on the most-updated Yonsei-Yale ($Y^2$) isochrones and HB evolutionary tracks with enhanced helium \citep{han2009,lee2015}, all constructed under the assumption that [$\alpha$/Fe] = 0.3. The \citet{reimers1977} mass-loss parameter $\eta$ was adopted to be 0.50, which best reproduces the observed HB morphology 
   of the old inner halo GCs with different metallicities 
   \citep[see also \citet{mcd2015}]{jang2015} 
  Following \citet{jang2014a}, mass dispersion on the HB was adopted to be $\sigma_{M}$ = 0.010 $M_{\sun}$.
  In the construction of synthetic CMDs from zero-age main-sequence (MS) to red giant branch (RGB) tip
  , we adopt the \citet{salpeter1955} initial mass function (IMF) with the standard exponent $\alpha$ = 2.3\footnote{We verified that the choice of $\alpha$ has negligible effect on the UV integrated photometry.}. 
 We assumed two age values of 12.0 and 12.5 Gyr and a grid of iron abundances that range from [Fe/H]=$-$2.5 to $-0.2$ in steps of 0.1 dex. 
  The stochastic effect stemming from a finite number of stars have been minimized by deriving mean values of 1,000 simulations, each with the sample size of $\sim$1,000 HB stars, at given parameter combination. 

\begin{deluxetable*}{ccccccccccc}
\tablenum{1}
\tablecaption{Integrated $IC_{\rm F275W,F336W,F438W}$ and $IC_{\rm F275W-F814W}$ colors of 1G and 2G stars in eleven GCs.\label{tab1}}
\tablewidth{10pt}
\setlength{\tabcolsep}{6pt}
\tablehead{
\colhead{NAME}&  \multicolumn{5}{c}{$IC_{\rm F275W,F336W,F438W}$} &\multicolumn{5}{c}{$IC_{\rm F275W-F814W}$} \\
\cline{3-5}
\cline{8-10}
\colhead{} &&\colhead{1G} & & \colhead{2G} &&& \colhead{1G} & & \colhead{2G} & }
\startdata
NGC 104 && 1.171$\pm$0.008 && 1.046$\pm$0.005 &&& 3.412$\pm$0.074 && 3.479$\pm$0.024 & \\
NGC 6121 && 0.980$\pm$0.040 && 0.743$\pm$0.017 &&& 2.561$\pm$0.154 && 2.588$\pm$0.077 & \\
NGC 6352 && 1.258$\pm$0.021 && 1.108$\pm$0.024 &&& 3.635$\pm$0.114 && 3.570$\pm$0.119 & \\
NGC 6362 && 1.128$\pm$0.023 &&0.893$\pm$0.047  &&& 2.810$\pm$0.123 && 2.619$\pm$0.210 & \\
NGC 6366 && 1.153$\pm$0.015 &&1.033$\pm$0.012  &&& 3.266$\pm$0.057 && 3.056$\pm$0.054 & \\
NGC 6388 && 1.414$\pm$0.029 &&1.089$\pm$0.009  &&& 4.001$\pm$0.077 && 3.478$\pm$0.022 & \\
NGC 6496 && 1.238$\pm$0.030 &&1.181$\pm$0.052  &&& 3.862$\pm$0.077 && 3.586$\pm$0.101 & \\
NGC 6624 && 1.246$\pm$0.015 &&1.040$\pm$0.010  &&& 3.241$\pm$0.102 && 3.666$\pm$0.051 & \\
NGC 6637 && 1.229$\pm$0.014 &&1.076$\pm$0.014  &&& 3.238$\pm$0.076 && 3.238$\pm$0.070 & \\
NGC 6652 && 1.232$\pm$0.020 &&1.101$\pm$0.012  &&& 3.096$\pm$0.143 && 3.294$\pm$0.082 & \\
NGC 6838 && 1.199$\pm$0.042 &&0.919$\pm$0.073  &&& 3.302$\pm$0.194 && 3.286$\pm$0.235 & \\
\enddata
\vspace{-30pt}
\end{deluxetable*}

To account for the effect of light-element variations on the magnitudes of 1G and 2G stars, we extracted the atmospheric parameters from each isochrone and HB evolutionary track point. These quantities are used to calculate a pair of synthetic spectra based on stellar models with the same values of effective temperature and gravity, but with different chemical composition. We derived a reference spectrum with solar content of C and N and with enhanced O abundance ([O/Fe]=0.4) and a comparison spectrum with C, N and O abundances representative of 2G stars of GCs [C/Fe]=$-0.50$, [O/Fe]=$-0.50$ and [N/Fe]=$1.21$ \citep[see][for details]{milone2017a}.   
Each spectrum has been derived by using ATLAS\,12 and SYNTHE programs \citep[e.g.][]{kurucz1970a, kurucz1981a, kurucz1993a, sbordone2004a, sbordone2007a, castelli2005a} and spans the wavelength interval from 2,000$\AA$ and 13,000$\AA$. 
  
The spectra have been convolved through the bandpasses of the ACS/WFC and WFC3/UVIS filters used in this paper to derive the corresponding magnitudes. Finally,  the  magnitude  differences  between  the  comparison  and  the  reference  spectrum have been added to the magnitudes of isochrone and simulated HB stars to account for the appropriate chemical composition of 2G stars.   
  
  We emphasize that the simulated photometry is used to qualitatively investigate the impact of multiple stellar populations on the integrated colors of GCs. A quantitative comparison between simulated and observed colors is beyond the purposes of this paper and should account for the specific properties of each cluster, including the number of populations, their chemical composition and, the fraction of stars in each of them.  
 
\begin{figure*}
\begin{center} 
  \includegraphics[height=13.0cm,trim={0cm 0cm 0cm 0cm},clip]{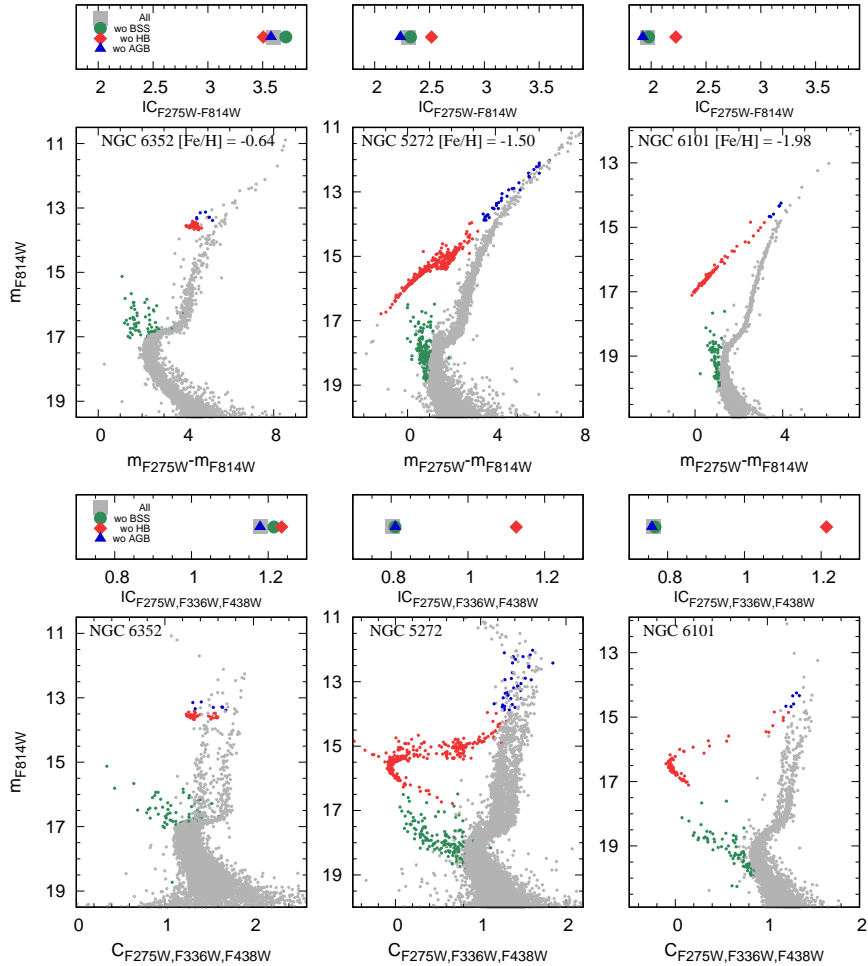}
  \caption{$m_{\rm F814W}$ vs.\,$m_{\rm F275W}-m_{\rm F814W}$ and \,$C_{\rm F275W,F336W,F438W}$ diagrams for three GCs with different metallicities, namely, NGC\,6352 (left), NGC\,5272 (middle) and NGC\,6101 (right). BSSs, HB stars and AGB stars are colored green, red and blue, respectively. 
  The average values of $IC_{\rm F275W-F814W}$ and $IC_{\rm F275W,F336W,F438W}$ for all cluster members and for all stars but BSSs, HB or AGB stars, are represented in the small panels above each CMD with gray squares, green circles, red diamonds, and blue triangles, respectively.}
   \label{fig:BSSs}
 \end{center} 
\end{figure*} 

\section{Integrated photometry of first- and second-generation stars}\label{sec:3}
In this section, we investigate the integrated photometry of the distinct stellar populations for a subsample of GCs to understand how each population affects the integrated color of each cluster.
We selected eleven GCs as ideal targets to derive integrated fluxes of 1G and 2G stars. Indeed, these are the only GCs where 1G and 2G define distinct sequences in their CMDs that can be unambiguously followed continuously along the main evolutionary stages of stellar evolution, from the MS to the the RGB and even along the AGB and the HB \citep[see][for details]{milone2012b, lagioia2021a, dondoglio2020a}.

In addition, we study the effect of blue straggler (BSS), HB and asymptotic giant branch (AGB) stars on the integrated colors.

\subsection{Integrated photometry of first- and second-generation stars}
To derive integrated photometry of the distinct stellar populations of GCs, we first identified the distinct groups of 1G and 2G stars along the MS, subgiant branch (SGB), red giant branch (RGB) and HB by following the procedure illustrated in Figure \ref{fig:NGC6637} for NGC\,6637. We used two different diagrams to disentangle the bulk of 1G and 2G stars among stars with different luminosities, which are represented with red and blue colors, respectively, in Figure \ref{fig:NGC6637}a. 

\begin{itemize}
    \item The $\Delta_{\rm {\it C} F275W,F336W,F438W}$ vs.\,$\Delta_{\rm F275W,F814W}$ chromosome map (ChM) is a pseudo two color diagram to maximize the separation among multiple MSs and RGBs \citep[][]{milone2015a}. As shown in Figure \ref{fig:NGC6637}d,e for NGC\,6637, 1G stars define the sequence around the origin of the ChM, whereas 2G stars are distributed towards high absolute values of $\Delta_{\rm {\it C} F275W,F336W,F438W}$ and $\Delta_{\rm F275W,F814W}$. 
    
    \item The ChM is also an efficient tool to separate 1G from 2G AGB stars \citep[][]{marino2017a}. We adopted the identification of 1G and 2G stars derived by \citet{lagioia2021a} from the ChM. 
    
    \item The $m_{\rm F336W}-m_{\rm F438W}$ vs.\,$m_{\rm F275W}-m_{\rm F336W}$ two color diagram is used to disentangle 1G and 2G stars along the SGB \citep[][]{milone2012a}, where 1G stars define the sequence of stars with redder $m_{\rm F275W}-m_{\rm F336W}$ colors
    (\ref{fig:NGC6637}b).
    
    \item The identification of 1G and 2G stars along the HB of metal rich clusters ([Fe/H]$\gtrsim -1.0$) is provided by \citet{dondoglio2020a} based on the $m_{\rm F336W}-m_{\rm F438W}$ vs.\,$m_{\rm F275W}-m_{\rm F336W}$ two-color diagram (\ref{fig:NGC6637}c).
    In the other analyzed GCs with blue HB we separate the bulk of 1G stars from the 2G based on distribution of simulated 1G stars by \citet{tailo2020a}.  
    
    \item To date, it is not possible to disentangle MPs among BSSs of GCs from photometry. Hence, we randomly associated BSSs to the 1G and the 2G. 
    To do this, we considered 1G members all BSSs where a number, randomly generated between 0 and 1 is smaller than the fraction 1G stars derived by \citet{milone2017a}. We applied the same method for the other stars, mostly binaries and faint MS-stars, that are not clearly separated into 1G and 2G. As we further discuss in the next sub-section, these stars provide a modest contribution to the integrated colors of GCs.

\end{itemize}

\begin{figure*} 
\begin{center} 
\includegraphics[height=6.2cm,trim={0cm 0cm 0cm 0cm},clip]{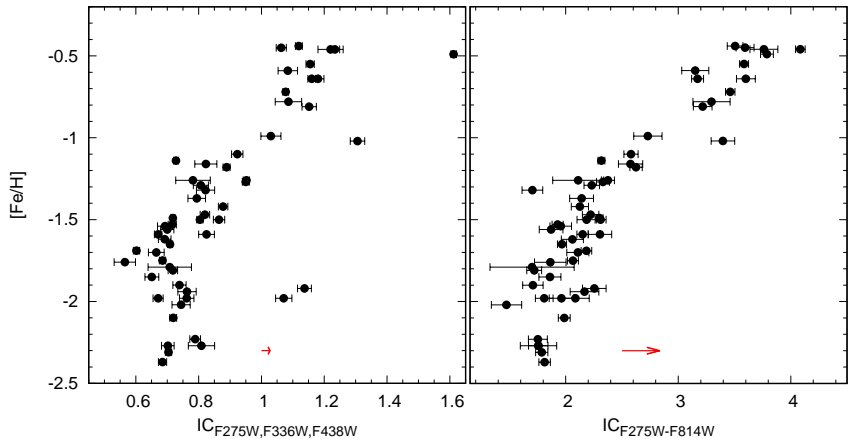}
  \caption{$IC_{\rm F275W,F336W,F438W}$ (left) and $IC_{\rm F275W-F814W}$ against metallicity (right) for the studied 56 Galactic GCs. The red arrows indicate the color shift corresponding to a reddening variation of E(B$-$V)=0.1 mag.}
  \label{fig:CvsFe}
 \end{center} 
\end{figure*} 

The 1G and 2G stars in eleven GCs, namely NGC\,104, NGC\,6121, NGC\,6352, NGC\,6362, NGC\,6366, NGC\,6388, NGC\,6496, NGC\,6624, NGC\,6637, NGC\,6652,  and NGC\,6838, have been identified by following the procedure above. These are the only GCs where the 1G and 2G stars can be clearly identified along most evolutionary stages. 
 We then estimated the integrated magnitudes and the corresponding uncertainties.
 The latter are derived by bootstrapping, with replacements over the sample of stars in each population. The derived errors refer to standard deviation of 1,000 bootstrapped measurements.
The resulting integrated colors, $IC_{\rm F275W,F336W,F438W}$ and $IC_{\rm F275W-F814W}$, are listed in Table \ref{tab1}.

We find that 1G stars of all the eleven GCs have redder integrated pseudo-colors than the 2G, with an average difference in $IC_{\rm F275W,F336W,F438W}$ of $\sim$0.2 mag.
On the other hand, there is no significant difference between 1G and 2G in $IC_{\rm F275W-F814W}$.


\begin{figure*} 
\begin{center} 
\includegraphics[height=6.2cm,trim={0.0cm 0.0cm 0cm 0cm},clip]{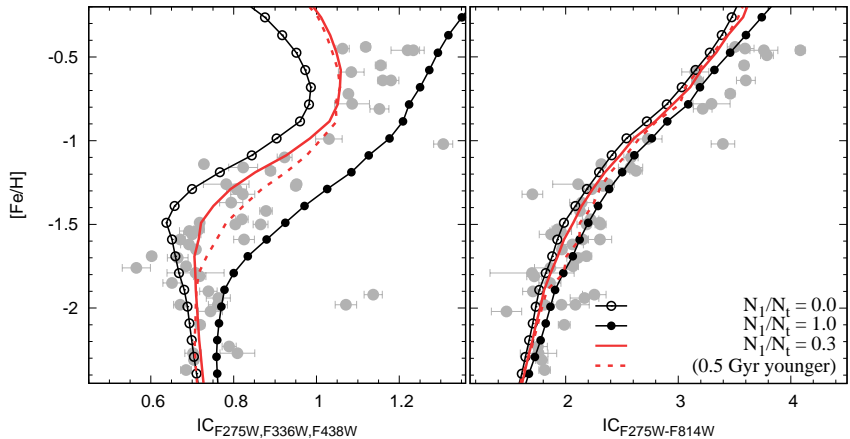}
  \caption{Same as Figure \ref{fig:CvsFe}, but with the integrated colors derived from simulated photometry superimposed on the observed points (gray dots). 
   The black filled and open dots correspond to 12.5 Gyr old  clusters composed of 1G and 2G stars alone, while red solid and dotted lines mark simulated GCs with a fraction of 1G stars, $N_{\rm 1}/N_{\rm t}$ =0.3 and ages of 12.5 Gyr and 12.0 Gyr, respectively. See text for details.   
  }
  \label{fig:CvsFewM}
 \end{center} 
\end{figure*} 

\subsection{The contribution of blue stragglers, horizontal branch, and asymptotic giant branch stars on the integrated colors}

The top and bottom panels of Figure \ref{fig:BSSs} show the $m_{\rm F814W}$ vs. $m_{\rm F275W}-m_{\rm F814W}$ and $C_{\rm F275W,F336W,F438W}$ diagrams, respectively, of three clusters with different metallicities and HB morphologies. These GCs are used as test cases to estimate the contribution of BSS, HB and AGB stars to the integrated colors the $IC_{\rm F275W,F336W,F438W}$ and $IC_{\rm F275W-F814W}$ of GCs. 
 Specifically, we selected a metal-rich GC with the red HB alone, NGC\,6352 ([Fe/H]=$-$0.64), 
  NGC\,5272 ([Fe/H]=$-$1.50), which is a metal-intermediate GC whose HB is well populated on both sides of the RR Lyrae instability strip and the metal-poor, blue-HB GC NGC\,6101 ([Fe/H]=$-$1.98) \citep[][2010 version]{harris1996a}.
  
To evaluate the effect of BSSs, we calculated the integrated values of $C_{\rm F275W,F336W,F438W}$ and $m_{\rm F275W-F814W}$ of each cluster by excluding the sample of BSSs that we identified by eye from the $m_{\rm F814W}$ vs.\,$m_{\rm F606W}-m_{\rm F814W}$ CMD (green points). Similarly, we estimated the contribution from HB and AGB stars (represented with red and blue points in the CMDs of Figure \ref{fig:BSSs}, respectively). 
The integrated $m_{\rm F275W}-m_{\rm F814W}$ and $C_{\rm F275W,F336W,F438W}$ colors derived by removing the group of BSSs, HB and AGB stars are presented in the small panels above the corresponding diagrams of Figure \ref{fig:BSSs}. 
The integrated colors strongly depend on metallicity, and range from $\sim$3.6 to $\sim$2.0 in $IC_{\rm F275W-F814W}$, and from $\sim$1.2 to $\sim$0.8 in $IC_{\rm F275W,F336W,F438W}$.

We find that HB stars have a small impact in the integrated colors of the metal-rich GC, but largely affect the $IC_{\rm F275W,F336W,F438W}$ and $IC_{\rm F275W-F814W}$ values of the two metal-poor clusters. 
The influence of AGB stars appears to be hardly revealed both on the $IC_{\rm F275W-F814W}$ and $IC_{\rm F275W,F336W,F438W}$ for the three GCs.

Similarly, BSSs have negligible effects of less than 0.01 mag on the integrated $C_{\rm F275W,F336W,F438W}$ and $IC_{\rm F275W-F814W}$ colors of NGC\,5272 and NGC\,6101.  They provide a slightly larger contribution  of the integrated colors of NGC\,6352, which ranges from $IC_{\rm F275W,F336W,F438W}$=1.18 to 1.21 and $IC_{\rm F275W-F814W}$=3.60 to 3.71, when BSSs are not taken into account.
This fact suggests that results on the integrated colors of 1G and 2G stars are poorly affected by BSSs. Similar conclusions are inferred from the remaining GCs. 
\section{The integrated colors of 56 Galactic Globular Clusters}\label{sec:4}

Based on the integrated magnitudes listed in Table \ref{56GCs}, we derived the $IC_{\rm F275W,F336W,F438W}$ and $IC_{\rm F275W-F814W}$ integrated colors for the 56 studied GCs. We find that the values of both $IC_{\rm F275W,F336W,F438W}$ and $IC_{\rm F275W-F814W}$  dramatically change from one cluster to another and span wide intervals of $\sim 0.9$ and $\sim 2.5$ mag, respectively.
In the following subsections, we analyze the relations between the integrated colors and the main properties of the host GCs and of their populations.

\subsection{Relations with cluster metallicity}
As shown in the left panel of Figure \ref{fig:CvsFe}, there is a strong correlation between the integrated $C_{\rm F275W,F336W,F438W}$ pseudo-color and the iron abundance of the host GC, \citep[from the 2010 version of the][catalog]{harris1996a} with the most metal-rich GCs having the reddest integrated colors. Similarly, $IC_{\rm F275W-F814W}$ correlates with [Fe/H] (right panel).
 An exception is provided by the bulk of metal-poor GCs with [Fe/H]$\lesssim -1.5$ in the left-panel plot, where most GCs exhibit similar values of $IC_{\rm F275W,F336W,F438W}$ regardless of [Fe/H].

To qualitatively investigate the behaviour of $IC_{\rm F275W,F336W,F438W}$ and $IC_{\rm F275W-F814W}$ as a function of [Fe/H], we simulated the same integrated colors for GCs with different fractions of 1G and 2G stars at different metallicities. Specifically, as shown in Figure \ref{fig:CvsFewM}, we simulated star clusters composed of 1G stars and 2G stars alone (black filled and open dots, respectively), and GCs with the prevalence of 2G stars ($N_{\rm 1}/N_{\rm t}$ = 0.3, red solid line) as observed in the majority of bright and massive Galactic GCs \citep[e.g.][]{milone2017a, dondoglio2020a}.

The simulated $IC_{F275W,F336W,F438W}$ color sequences, shown in the left panel of Figure \ref{fig:CvsFewM}, show that clusters containing only 2G stars have bluer integrated colors and populate the leftmost portion of the diagram, while clusters populated by 1G stars only have redder $IC_{F275W,F336W,F438W}$ at a given metallicity. On the other hand, the $IC_{F275W-F814W}$ color differences among the simulated clusters seems less evident.

To estimate the effect of age on the scatter observed among GCs at a given metalicity, we simulated GCs with $N_{\rm 1}/N_{\rm t}$ = 0.3, but with 0.5 Gyr younger (red dotted line). The corresponding theoretical $IC_{F275W,F336W,F438W}$ and $IC_{F275W-F814W}$ sequences in the plot indicate that the effect of age is negligible at metal-poor and metal-rich regime, while it has a bigger impact at intermediate metallicities, $-1.7\lesssim$[Fe/H]$\lesssim -1.0$.

\begin{figure*} 
\begin{center} 
  \includegraphics[height=6.5cm,trim={0cm 0cm 0cm 0cm},clip]{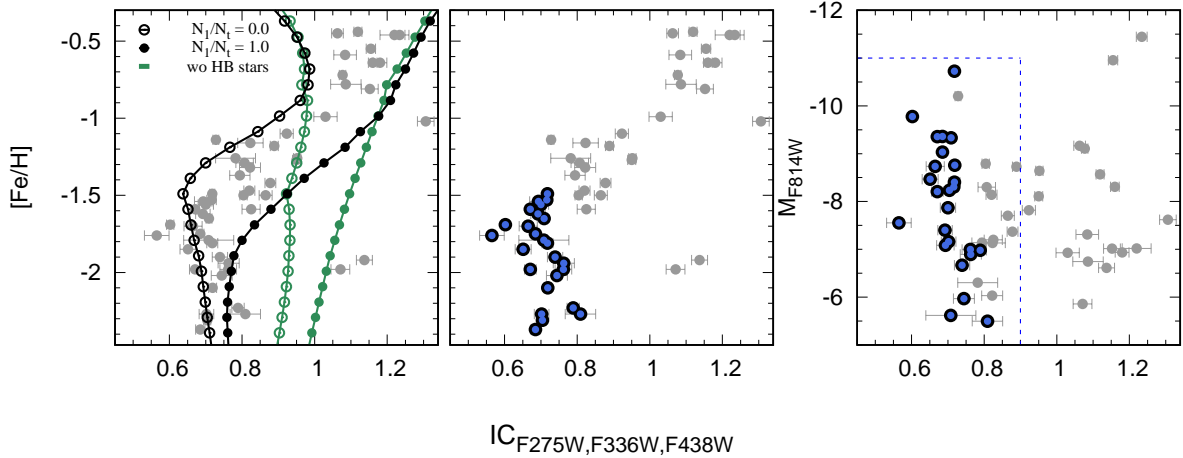}
  \caption{\textit{Left panel.} Comparison of the observed MW GCs shown as gray with models with and without HB stars (black and green lines, respectively) in $IC_{\rm F275W,F336W,F438W}$ vs.\,[Fe/H] diagram.
  The same diagram is reproduced in the middle panel, where we marked with colored dots the metal-poor GCs with nearly constant $IC_{\rm F275W,F336W,F438W}$ values. Right panel shows the $M_{\rm F814W}$ vs.\,$IC_{\rm F275W,F336W,F438W}$ for all GCs. }
    \label{fig:Cmetalpoor}
 \end{center} 
\end{figure*} 

\subsection{Integrated colors and multiple populations in metal-poor globular clusters}\label{sec:metalpoor}

As illustrated in the left panel of Figure \ref{fig:CvsFe}, most metal-poor GCs with [Fe/H]$\lesssim -1.5$ have nearly constant $IC_{\rm F275W,F336W,F438W} \sim 0.7$, while in the remaining clusters $IC_{\rm F275W,F336W,F438W}$ defines a strong correlation with cluster metallicity. NGC\,2298 ([Fe/H] = $-$1.92) and NGC\,5466 ([Fe/H] = $-$1.98), which exhibit larger values of $IC_{\rm F275W,F336W,F438W}$ than the bulk of metal-poor GCs, are remarkable exceptions. Admittedly, we do not have an explanation for the red $IC_{\rm F275W,F336W,F438W}$ colors of these two clusters. Possibly, this is not an effect due to to multiple populations as, the fraction of 2G stars and the maximum variation of helium and nitrogen of both NGC\,2298 and NGC\,5466, is comparable with that of the remaining GCs with similar metallicity and masses. 

To investigate the relation between $IC_{\rm F275W,F336W,F438W}$ 
 and the multiple-population phenomenon, in the middle panel of Figure \ref{fig:Cmetalpoor} we mark with colored dots 26 GCs with  $IC_{\rm F275W,F336W,F438W} \lesssim 0.8$ that lie on a vertical sequence in the $IC_{\rm F275W,F336W,F438W}$ vs.\,[Fe/H] diagram.

The same clusters are highlighted with the same symbols in the $M_{\rm F814W}$ vs.\,$IC_{\rm F275W,F336W,F438W}$ CMD plotted in the right panel of Figure \ref{fig:Cmetalpoor}  where they define a vertical sequence on the blue side of the CMD.

Intriguingly, the selected clusters are all metal-poor GCs with HB ratio, HBR$>0.5$\footnote{The HB type is defined as (B$–$R)/(B$+$V$+$R), where B, V, and R are the numbers of blue HB, RR Lyrae variables, and red HB stars, respectively \citep{lee1994}.}. 
The fact that the HBs of these clusters are dominated by blue-HB stars, is one of the reasons for the poor dependence of their $IC_{\rm F275W,F336W,F438W}$ colors from cluster metallicity. 
Indeed, blue-HB stars of metal-poor GCs never exceed the blue threshold of $C_{\rm F275W,F336W,F438W} \sim -0.2$ (see e.g. Figure \ref{fig:BSSs}). Since HB stars provide a major contribution to the integrated color of metal-poor GCs, the dominance of blue HB stars makes the $IC_{\rm F275W,F336W,F438W}$ color poorly sensitive to the overall GC metallicity.  This fact is consistent with the simulated colors plotted in the left panel of Figure \ref{fig:CvsFewM}.  As shown in the left panel of Figure \ref{fig:Cmetalpoor}, the integrated $IC_{\rm F275W,F336W,F438W}$  colors calculated for GCs without the HB (green lines) correlate with [Fe/H], but this correlation disappears among metal-poor GCs when we account for the contribution of HB stars (black lines).

\begin{figure*} 
\begin{center} 
  \includegraphics[height=8.5cm,trim={0cm 0.0cm 0cm 0cm},clip]{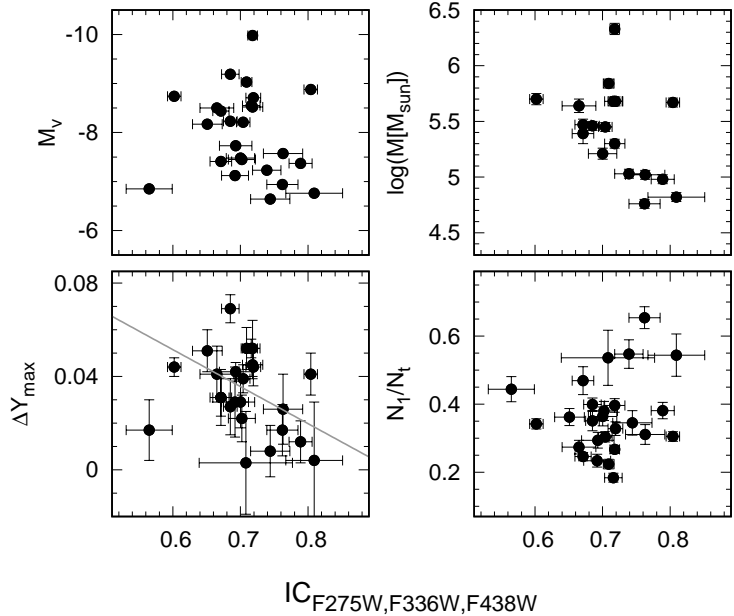}
  \caption{$IC_{\rm F275W,F336W,F438W}$ of the metal-poor GCs selected in Figure \ref{fig:Cmetalpoor} as a function of the absolute magnitudes, GC masses, fractions of 1G stars and maximum internal helium variations. }
  \label{fig:Cmetalpoor3}
 \end{center} 
\end{figure*} 

Having showed that the $IC_{\rm F275W,F336W,F438W}$ color of metal-poor GCs is poorly affected by cluster metallicity, we investigate the physical reasons of the $IC_{\rm F275W,F336W,F438W}$  scatter observed among metal-poor GCs. 
Recent works have shown that both the fraction of 2G stars and the maximum internal helium variation anti-correlate with the absolute magnitude of the host GC \citep[hence, correlate with cluster mass][]{milone2017a}.

The scarce dependence of metallicity on $IC_{\rm F275W,F336W,F438W}$ among the selected 26 metal-poor GCs allows us to directly compare their pseudo colors with those parameters that are  related with the multiple-population phenomenon, including the GC absolute luminosity, mass, fraction of 1G stars and maximum internal variation of helium \citep[from][]{harris1996a, milone2017a, milone2018a}.

Figure \ref{fig:Cmetalpoor3} reveals that the GCs with bluer $IC_{\rm F275W,F336W,F438W}$ exhibit brighter cluster luminosity, higher mass and  $\Delta Y_{\rm max}$, 
 and lower fractions of 1G stars $N_{\rm 1}/N_{\rm t}$.  
 GCs with integrated colors bluer than the average value,  $IC_{\rm F275W,F336W,F438W}=0.73$ have, on average, higher helium mass fractions ($\Delta Y=$0.020$\pm$0.007) and lower fractions of 1G stars ($\Delta N_{\rm 1}/N_{\rm t}=-0.10 \pm 0.06$).

\begin{figure*} 
\begin{center} 
  \includegraphics[height=9.0cm,trim={0.0cm 0cm 0.0cm 0cm},clip]{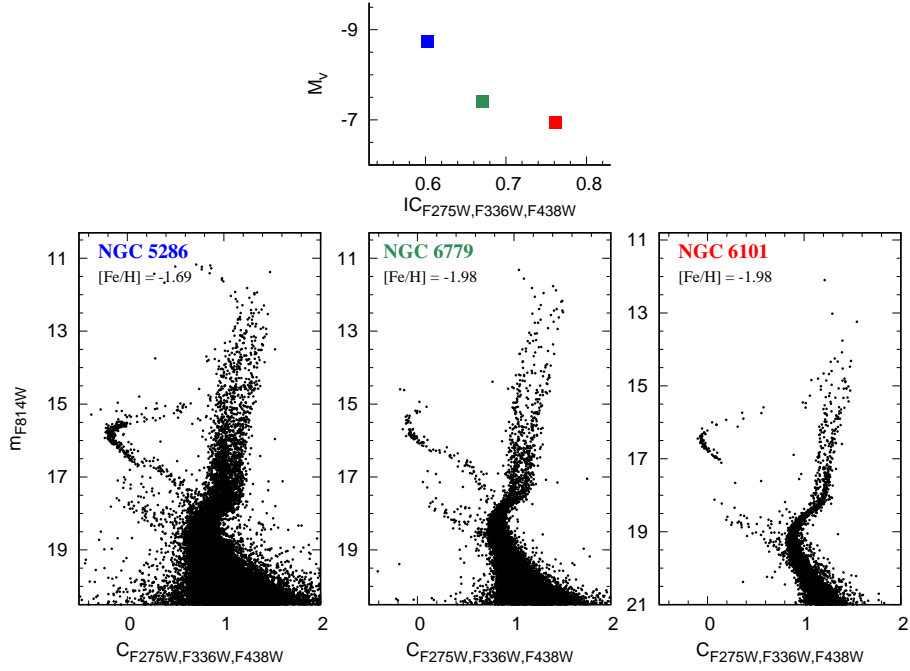}
  \caption{$m_{\rm F814W}$ versus $C_{\rm F275W,F336W,F438W}$ pseudo-CMDs of NGC\,5286, NGC\,6779, and NGC\,6101. The upper panel show the GC absolute magnitude against the integrated pseudo-color $IC_{\rm F275W,F336W,F438W}$.} 
            \label{fig:Cmetalpoor2}
 \end{center} 
\end{figure*}

  There are mild correlations between $IC_{\rm F275W,F336W,F438W}$ and $\Delta Y_{\rm max}$, where the least-square fit line is given by $\Delta Y_{\rm max} = -0.16 \cdot IC_{\rm F275W,F336W,F438W} + 0.15$. 
    As an example, Figure \ref{fig:Cmetalpoor2} compares the $m_{F814W}$ vs.\,$C_{\rm F275W,F336W,F438W}$ pseudo CMDs of three GCs out of them with different absolute magnitudes, NGC\,5286, NGC\,6779, and NGC\,6101. Clearly, NGC\,5286 and NGC\,6101, which are the most- and the least-massive of the three GCs, respectively, exhibits the bluest and the reddest values of $IC_{\rm F275W,F336W,F438W}$.
These findings suggest that the $IC_{\rm F275W,F336W,F438W}$ integrated color of metal-poor GCs is sensitive of the complexity of multiple populations in GCs.

Nevertheless, we notice large cluster-to-cluster scatter in the relations between the integrated color and the parameters. This fact demonstrates that, in addition to the quantities plotted on the ordinate of Figure \ref{fig:Cmetalpoor3},  one or more second parameters affect the $IC_{\rm F275W,F336W,F438W}$ value of metal-poor GCs.

\section{Detection of signals of multiple populations from integrated light}\label{sec:5}

In this section, we investigate Galactic GCs in the $IC_{F275W,F336W,F438W}$ vs.\,$IC_{F275W,F814W}$ plane, which is built by using GC colors alone and does not require information on GC metallicity. 

Figure \ref{fig:TWOCOLOR} reveals that Galactic GCs define a strong correlation between the $IC_{F275W,F336W,F438W}$ and $IC_{F275W,F814W}$ integrated colors. Metal-poor clusters typically exhibit lower values of both $IC_{F275W,F336W,F438W}$ and $IC_{F275W,F814W}$, whereas metal-rich clusters have higher $IC_{F275W,F336W,F438W}$ and $IC_{F275W,F814W}$ colors.
Although GCs are distributed along a distinct sequence, 
 we notice a significant $IC_{F275W,F336W,F438W}$ scatter among clusters with similar values of $IC_{F275W-F814W}$.

The same simulations introduced in Figure \ref{fig:CvsFewM} are superimposed on the upper-right panel of Figure \ref{fig:TWOCOLOR}, where the groups of simulated GCs define distinct sequences. Clusters composed of 2G stars alone ($N_{\rm 1}/N_{\rm t}$ = 0, open dots) have bluer $IC_{F275W,F336W,F438W}$ colors than the 1G clusters ($N_{\rm 1}/N_{\rm t}$ = 1, filled dots) with the same metallicity. 
The simulated GCs that host both 1G and 2G stars ($N_{\rm 1}/N_{\rm t}$ = 0.3, red lines) occupy intermediate positions.
Noticeably, the sequences of GCs with ages of 12.5 and 12.0 Gyr (red solid and dot lines) are almost overimposed on each other, thus demonstrating that age differences poorly affect the position of GCs in the $IC_{F275W,F336W,F438W}$ vs.\,$IC_{F275W-F814W}$ plane.

Although a quantitative comparison between the simulated colors and the integrated colors of Galactic GCs is beyond purposes of this work, our results demonstrate that the $IC_{F275W,F336W,F438W}$ vs.\,$IC_{F275W,F814W}$ diagram 
is an efficient tool to detect signals of multiple populations.

This conclusion is well corroborated by the results on eleven clusters listed in Table \ref{tab1}. As shown in the bottom-right panel of Figure \ref{fig:TWOCOLOR}, their 1G stars (red triangles) exhibit  redder $IC_{F275W,F336W,F438W}$ colors than the 2G (blue squares), with the same $IC_{F275W,F814W}$.

To further investigate the relation between integrated colors and multiple stellar populations, we used pink colors to mark clusters with $M_{\rm V}>-7.3$ in the bottom-left panel of Figure \ref{fig:TWOCOLOR}.  The evidence that these low-mass clusters exhibit, on average, higher values of  $IC_{F275W,F336W,F438W}$  than GCs with similar $IC_{F275W,F814W}$ color, further proves that the position of a GC in the two-color diagram of Figure \ref{fig:TWOCOLOR} depends on its multiple stellar populations.

\subsection{Correlations with the properties of multiple populations}
\begin{figure*}
\begin{center} 
  \includegraphics[height=8.5cm,trim={0cm 0cm 0cm 0cm},clip]{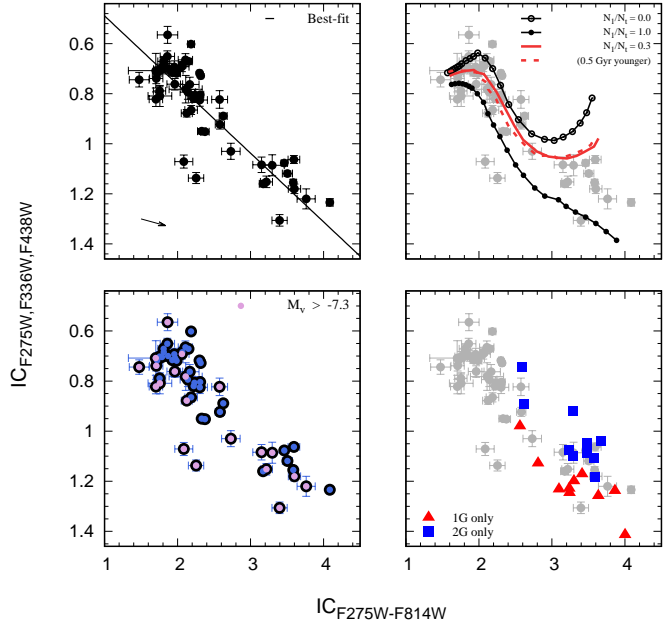}
  \caption{
  $IC_{F275W,F336W,F438W}$ vs.\,$IC_{F275W-F814W}$ integrated two-color diagrams for MW GCs.
  In the upper-left panel, we superimposed on the observed GCs the fiducial line. 
  The arrow indicates the reddening vector corresponding to a reddening variation of E(B$-$V)=0.1. 
  In the upper-right panel, the red lines mark the integrated colors of simulated GCs that comprises the 30\%, and the simulated clusters composed of 1G and 2G alone (black filled and open dots, respectively). 
  In bottom-left panel, the GCs with absolute magnitude $M_{\rm v}$ $> -7.3$ are marked with pink dots, whereas the remaining clusters are colored blue.  Red triangles and blue squares in the bottom-right panel indicate 1G and 2G stars, respectively, of eleven GCs listed in Table \ref{tab1}. See text for details.
 } 
  \label{fig:TWOCOLOR}
\end{center} 
\end{figure*}

To remove the effect of GC metallicity,
 we derived the best-fit straight line (black line in the top-left panel of Figure \ref{fig:TWOCOLOR}) and the color residuals, $\Delta IC_{F275W,F336W,F438W}$, which are the $IC_{F275W,F336W,F438W}$ distances of each cluster from the line calculated at the same $IC_{F275W,F814W}$ value.

\begin{figure*} 
\begin{center} 
  \includegraphics[height=6.7cm,trim={0.0cm 0cm 0cm 9.5cm},clip]{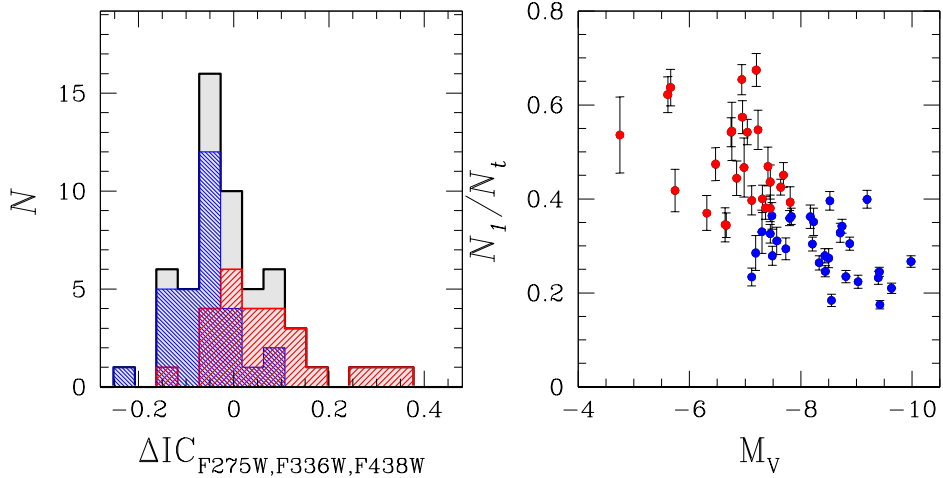}
  \caption{\textit{Left.} Histogram distribution of the   $IC_{F275W,F336W,F438W}$ residuals. \textit{Right.} Fraction of 1G stars as a function of the absolute magnitude of the host GC. The $\Delta IC_{F275W,F336W,F438W}$ histogram distribution of the groups of red and blue GCs selected in the right-panel plot are represented with the same colors in the right panel.}
      \label{fig:histo}
 \end{center} 
\end{figure*}

To investigate the relation between color residuals and multiple populations, we plot  the histogram distribution of $\Delta IC_{F275W,F336W,F438W}$ in the left panel of Figure \ref{fig:histo} (gray shaded histogram).  We then defined two groups of clusters based on their fractions of 1G stars and absolute luminosity as shown in the right panel of Figure \ref{fig:histo}. Clusters with high absolute luminosity and low fractions of 1G stars are colored blue. This sample hosts stellar populations with extreme properties such as large internal variations in helium and nitrogen. On the contrary, the remaining GCs, which are plotted with red symbols, exhibit  moderate light-element star-to-star variations \citep[][]{milone2018a, milone2020a}.

The selected groups of red and blue clusters show different distributions of $\Delta IC_{F275W,F336W,F438W}$ as illustrated in the left panel of Figure \ref{fig:histo}, where we  use the same colors to represent the corresponding histogram distributions.  
 Specifically, the sample of GCs with extreme chemical composition exhibit, on average smaller $\Delta IC_{F275W,F336W,F438W}$ values  than the clusters with moderate chemical variations. This difference of 0.12$\pm$0.02 mag between the two groups corroborates the evidence that $\Delta IC_{F275W,F336W,F438W}$ depends on the properties of multiple populations.

\begin{figure*} 
\begin{center} 
  \includegraphics[height=9.2cm,trim={0.0cm 0cm 0cm 0cm},clip]{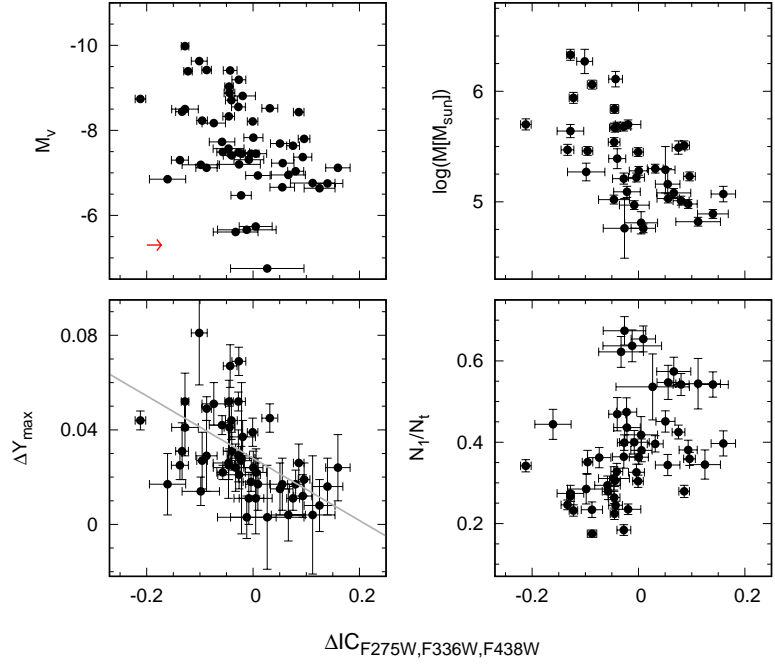}
  \caption{ The color residuals, $\Delta IC_{F275W,F336W,F438W}$, 
   are plotted against the absolute visual magnitude $M_{\rm v}$, cluster mass, maximum internal helium variation, $\Delta Y_{\rm max}$,  and fraction of 1G stars, $N_{\rm 1}/N{\rm t}$. 
  A best-fit line is provide by $\Delta Y_{\rm max}$ = $-$0.13 $\cdot \Delta IC_{F275W,F336W,F438W}$ $+$ 0.028 (see text for details).}
      \label{fig:residuals}
 \end{center} 
\end{figure*}

We plot in the bottom panels of Figure \ref{fig:residuals} the residuals, $\Delta IC_{F275W,F336W,F438W}$, against the maximum internal helium variation of GC stars, $\Delta Y_{\rm max}$, \citep[from][]{milone2018a} and the fraction of 1G stars $N_{1}/N_{\rm t}$ \citep[from][]{milone2017a, dondoglio2020a}. 
We find that $\Delta IC_{F275W,F336W,F438W}$ depends on $\Delta Y_{\rm max}$ and $N_{1}/N_{\rm t}$. In particular, GCs having negative values of $\Delta IC_{F275W,F336W,F438W}$ exhibit, on average, larger internal helium variations (by $\Delta Y=0.021 \pm 0.05$)
 and smaller fractions of 1G stars ($\Delta N_{1}/N_{\rm tot}=-0.11 \pm 0.03$) than GCs with $\Delta IC_{F275W,F336W,F438W}>0$. 
 
We have least-squares fitted the $\Delta Y_{\rm max}$ versus $\Delta IC_{F275W,F336W,F438W}$ with a straight line, as shown in the bottom-left panel of Figure ~\ref{fig:residuals}, where the best-fitting line is given by $\Delta Y_{\rm max}$ = -0.13$\cdot \Delta IC_{F275W,F336W,F438W}$ + 0.028. 
Noticeably, we observe a significant scatter with  respect to the best-fit line, which is wider than what we expect from observational errors alone. 
 This fact indicates that, in addition to helium variation, one or more second parameters affect the $\Delta IC_{F275W,F336W,F438W}$ quantity.
 Although a detailed investigation of these second parameters is beyond the purposes of this work, we suggest that the number of sub-populations within the 1G and the 2G, the relative number of stars and the specific distribution of He, N, and of other elements, are likely responsible for the observed scatter. In addition, as shown in Figure \ref{fig:TWOCOLOR}, age differences can affect the integrated $IC_{F275W,F336W,F438W}$ color. Similar discussion can be extended to all panels of Figure~\ref{fig:residuals}.
 
We also observe that the $\Delta IC_{F275W,F336W,F438W}$ decreases with the mass of a GC and increases with $M_{\rm V}$. This fact is quite expected given that the helium abundance of GCs correlates with GC mass \citep[][]{milone2018a}. Clearly, the fact that high mass GCs tend to have larger fraction of 2G stars as well as higher variations of helium and nitrogen, makes it challenging to disentangle the contribution of each of the parameter on the GC integrated light.

\section{Summary and conclusion}\label{sec:6}
We investigate 56 Galactic GCs where multiple populations have been widely studied and characterized by means of multi-band photometry of resolved stars \citep[e.g.][]{piotto2015a, milone2017a, tailo2020a, lagioia2021a}.
 The main goal is testing whether the presence of populations can be inferred from integrated photometry. 

Specifically, we exploited the photometric catalogs by \citet{nardiello2018a}, which are based on images collected through the F275W, F336W, F438W of UVIS/WFC3 and the F814W of WFC/ACS on board {\it HST} \citep[][]{sarajedini2007a, piotto2015a}. Results are based on the integrated color $IC_{F275W-F814W}$ and the integrated pseudo color $IC_{F275W,F336W,F438W}$ of GC stars. 

The main findings can be summarized as follows.

\begin{itemize}

    \item The metallicity of the host GCs exerts the strongest influence on both integrated colors. $IC_{\rm F275W,F336W,F438W}$ and $IC_{\rm F275W-F814W}$ span wide color ranges of $\sim$0.9 and  $\sim$2.5 mag, respectively, and correlate with cluster metallicity.

    \item In the eleven GCs, 
     where the 1G and 2G stars can be distinguished along most evolutionary stages, we derived the integrated colors of each stellar population. We find that 1G and 2G stars share similar values of $IC_{\rm F275W-F814W}$, while 1G stars have significantly redder $IC_{\rm F275W,F336W,F438W}$ colors than the 2G. 
    This fact demonstrates that, in addition to GC metallicity, the integrated pseudo color $IC_{\rm F275W,F336W,F438W}$ is sensitive to multiple stellar populations.

    \item The $IC_{\rm F275W,F336W,F438W}$ pseudo-colors of the metal-poor GCs ($[Fe/H] \lesssim -1.5$) 
     depend on the properties of multiple populations. Specifically, GCs with bluer $IC_{\rm F275W,F336W,F438W}$ colors are more luminous and massive, go through more severe helium enhancements, and exhibit higher number fractions of 2G stars.

    \item 
    The two-color diagram $IC_{\rm F275W,F336W,F438W}$ versus $IC_{\rm F275W-F814W}$ 
     is an efficient tool to detect signals of GC multiple populations. 
     Indeed, the position of each cluster in this plane mostly depends on cluster metallicity and on multiple population properties.
      After removing the dependence from [Fe/H], the residual  pseudo-colors $\Delta IC_{\rm F275W,F336W,F438W}$ 
      increase with the absolute magnitude and the number fraction of 1G stars 
      and decrease with the cluster mass and the maximum internal helium variation of GC stars. Hence, GCs with negative $\Delta IC_{\rm F275W,F336W,F438W}$ host multiple populations with more-extreme properties than the remaining GCs.
      Noticeably, we detect a large cluster-to-cluster scatter in the relations between the integrated-color residuals $N_{\rm 1}/N_{\rm t}$, $\Delta Y_{max}$, cluster luminosity and mass. This fact suggests that multiple populations alone may not entirely account for $\Delta IC_{\rm F275W,F336W,F438W}$ value of each cluster.

\end{itemize}

Nearly all photometric studies on multiple populations in GCs are based on resolved stars. 
The fact that it is not possible to disentangle distinct stars in very distant GCs has largely prevented the investigation of the multiple population phenomenon outside the local group. 

 Important effort to ascertain whether the multiple-population phenomenon also exists in the globulars of giant elliptical galaxies is provided by \citet{bellini2015} based on GC integrated light in the F275W, F606W and F814W filters of {\it HST}. These authors explored M\,87, the massive cD galaxy in the Virgo cluster, which hosts  $\sim$10,000 GCs.  
They identified a small group of globulars with a significant radial UV-optical color gradient, which represent the best candidate GCs hosting multiple stellar populations. Indeed, color gradient is consistent with multiple-population GCs, where the stellar populations with extreme helium content are more centrally concentrated than the 1G. 

Previous investigation by \citet{sohn2006} and \citet{kaviraj2007} have identified some M\,87 GCs that are red in optical and even near-UV bands but with extremely blue Far-UV colors. 
 Such color behaviour is attributed to extreme horizontal-branch (HB) stars in these red GCs, which are possibly associated with super helium-rich stellar populations. 
 Although these works provide important hints that the multiple-population phenomenon is not a peculiarity of GCs of the Milky Way and its satellites, their conclusions are based on a small number of candidate clusters with multiple populations.

Spectroscopy is another promising technique for hunting multiple populations in extragalactic GCs.
Studies based on the detailed abundance analysis from integrated light spectra of M\,31 GCs provide evidence for star-to-star abundance variations in Mg, Na, and Al together with correlations of abundance ratios of some Mg and Na with cluster luminosity and velocity dispersion \citep[][]{bastian2019,lardo2017,larsen2012,larsen2014b,larsen2018,sakari2015,schiavon2013}. Additional signals of multiple populations have been detected from integrated spectroscopy of clusters Large Magellanic Cloud, Fornax dwarf spheroidal, NGC 147, NGC 6822, NGC 1316, M33 and WLM dwarf galaxy \citep[][]{bastian2019,lardo2017,larsen2012,larsen2014b,larsen2018,sakari2015,schiavon2013}.

In this paper we have introduced a new diagram, based on integrated photometry in the F275W, F336W, F438W and F814W photometric bands of {\it HST} that allows to identify GCs with multiple populations. 
Our finding provides a new and efficient tool to extend the investigation of multiple populations outside the local group and understand whether the formation of GCs and of their stellar populations is universal or depends on the host galaxy.

\clearpage

\tablenum{2}
\startlongtable
\begin{deluxetable*}{lccccccc}
\centering
\tablewidth{0pt}
\setlength{\tabcolsep}{11pt}
\tablecaption{Integrated magnitudes and colors of Milky Way Globular Clusters\label{tab2}}
\tablehead{
ID &F275W&F336W&F438W&F606W&F814W&$IC_{\rm F275W,F336W,F438W}$&${IC_{\rm F275W-F814W}}$
}
\startdata
NGC 104 &        -5.643&   -6.953&   -7.187&   -8.288&   -9.105&    1.077$\pm$0.009&    3.462$\pm$0.042  \\
NGC 288 &        -4.329&   -4.953&   -4.754&   -5.466&   -6.033&    0.822$\pm$0.029&    1.704$\pm$0.092  \\
NGC 362 &        -6.264&   -7.135&   -7.054&   -7.964&   -8.639&    0.952$\pm$0.010&    2.375$\pm$0.058  \\
NGC 1261&        -5.778&   -6.642&   -6.556&   -7.446&   -8.109&    0.950$\pm$0.011&    2.331$\pm$0.058  \\
NGC 1851&        -6.104&   -7.004&   -7.015&   -8.010&   -8.729&    0.889$\pm$0.011&    2.625$\pm$0.055  \\
NGC 2298&        -4.631&   -5.589&   -5.427&   -6.116&   -6.691&    1.121$\pm$0.022&    2.060$\pm$0.105  \\
NGC 2808&        -8.144&   -8.768&   -8.620&   -9.541&  -10.205&    0.728$\pm$0.008&    2.317$\pm$0.030  \\
NGC 3201&        -4.896&   -5.690&   -5.658&   -6.546&   -7.200&    0.825$\pm$0.025&    2.304$\pm$0.105  \\
NGC 4590&        -5.227&   -5.843&   -5.670&   -6.388&   -6.980&    0.789$\pm$0.017&    1.753$\pm$0.084  \\
NGC 4833&        -6.605&   -7.126&   -6.995&   -7.816&   -8.466&    0.651$\pm$0.022&    1.860$\pm$0.098  \\
NGC 5024&        -6.772&   -7.387&   -7.282&   -8.108&   -8.760&    0.719$\pm$0.011&    1.987$\pm$0.053  \\
NGC 5053&        -3.738&   -4.345&   -4.143&   -4.889&   -5.496&    0.809$\pm$0.042&    1.758$\pm$0.162  \\
NGC 5272&        -6.478&   -7.241&   -7.200&   -8.099&   -8.789&    0.804$\pm$0.010&    2.310$\pm$0.049  \\
NGC 5286&        -7.592&   -8.189&   -8.184&   -9.109&   -9.776&    0.602$\pm$0.010&    2.184$\pm$0.047  \\
NGC 5466&        -3.866&   -4.776&   -4.622&   -5.313&   -5.885&    1.065$\pm$0.026&    2.019$\pm$0.123  \\
NGC 5897&        -4.959&   -5.531&   -5.364&   -6.071&   -6.667&    0.739$\pm$0.021&    1.708$\pm$0.091  \\
NGC 5904&        -6.068&   -6.815&   -6.755&   -7.635&   -8.299&    0.807$\pm$0.024&    2.231$\pm$0.069  \\
NGC 5927&        -5.640&   -7.157&   -7.062&   -8.673&   -9.428&    1.612$\pm$0.009&    3.788$\pm$0.057  \\
NGC 5986&        -7.208&   -7.828&   -7.777&   -8.685&   -9.358&    0.671$\pm$0.012&    2.150$\pm$0.051  \\
NGC 6093&        -6.970&   -7.540&   -7.425&   -8.356&   -9.034&    0.685$\pm$0.010&    2.064$\pm$0.049  \\
NGC 6101&        -5.035&   -5.701&   -5.605&   -6.363&   -6.997&    0.762$\pm$0.023&    1.962$\pm$0.108  \\
NGC 6121&        -4.559&   -5.432&   -5.483&   -6.431&   -7.134&    0.823$\pm$0.035&    2.576$\pm$0.108  \\
NGC 6144&        -5.690&   -6.148&   -6.042&   -6.873&   -7.553&    0.565$\pm$0.034&    1.863$\pm$0.140  \\
NGC 6171&        -4.221&   -5.670&   -5.813&   -6.885&   -7.619&    1.306$\pm$0.023&    3.397$\pm$0.105  \\
NGC 6205&        -6.377&   -6.959&   -6.825&   -7.660&   -8.304&    0.716$\pm$0.013&    1.928$\pm$0.049 \\
NGC 6218&        -4.998&   -5.689&   -5.587&   -6.487&   -7.139&    0.794$\pm$0.028&    2.141$\pm$0.106  \\
NGC 6254&        -5.998&   -6.551&   -6.405&   -7.238&   -7.868&    0.700$\pm$0.021&    1.870$\pm$0.105  \\
NGC 6304&        -5.571&   -6.926&   -7.217&   -8.367&   -9.165&    1.063$\pm$0.016&    3.594$\pm$0.080  \\
NGC 6341&        -6.446&   -6.986&   -6.822&   -7.606&   -8.233&    0.704$\pm$0.010&    1.787$\pm$0.054  \\
NGC 6352&        -3.335&   -4.777&   -5.039&   -6.181&   -6.936&    1.180$\pm$0.019&    3.601$\pm$0.084  \\
NGC 6362&        -4.196&   -5.236&   -5.245&   -6.208&   -6.927&    1.030$\pm$0.032&    2.730$\pm$0.126  \\
NGC 6366&        -4.159&   -5.388&   -5.532&   -6.657&   -7.310&    1.084$\pm$0.031&    3.151$\pm$0.040  \\
NGC 6388&        -7.371&   -8.738&   -8.950&  -10.170&  -10.956&    1.155$\pm$0.013&    3.585$\pm$0.135  \\
NGC 6397&        -4.497&   -5.009&   -4.778&   -5.426&   -5.970&    0.744$\pm$0.029&    1.474$\pm$0.043  \\
NGC 6441&        -7.330&   -8.982&   -9.340&  -10.625&  -11.444&    1.234$\pm$0.015&    4.085$\pm$0.124  \\
NGC 6496&        -3.257&   -4.710&   -4.943&   -6.132&   -7.018&    1.220$\pm$0.040&    3.761$\pm$0.357  \\
NGC 6535&        -3.916&   -4.478&   -4.332&   -5.039&   -5.617&    0.708$\pm$0.069&    1.701$\pm$0.066  \\
NGC 6541&        -6.683&   -7.221&   -7.041&   -7.805&   -8.403&    0.718$\pm$0.015&    1.720$\pm$0.086  \\
NGC 6584&        -5.517&   -6.306&   -6.229&   -7.066&   -7.703&    0.865$\pm$0.018&    2.186$\pm$0.070  \\
NGC 6624&        -5.064&   -6.407&   -6.632&   -7.771&   -8.569&    1.119$\pm$0.011&    3.505$\pm$0.053  \\
NGC 6637&        -5.135&   -6.410&   -6.526&   -7.580&   -8.307&    1.160$\pm$0.012&    3.172$\pm$0.083  \\
NGC 6652&        -3.798&   -5.074&   -5.198&   -6.274&   -7.014&    1.152$\pm$0.023&    3.217$\pm$0.098  \\
NGC 6656&        -6.628&   -7.195&   -7.097&   -8.031&   -8.736&    0.665$\pm$0.025&    2.108$\pm$0.098  \\
NGC 6681&        -5.342&   -5.939&   -5.844&   -6.705&   -7.400&    0.692$\pm$0.020&    2.059$\pm$0.031  \\
NGC 6715&        -8.424&   -9.088&   -9.034&  -10.029&  -10.725&    0.718$\pm$0.007&    2.301$\pm$0.227  \\
NGC 6717&        -4.193&   -4.894&   -4.815&   -5.632&   -6.304&    0.782$\pm$0.055&    2.111$\pm$0.063  \\
NGC 6723&        -5.239&   -6.170&   -6.179&   -7.117&   -7.819&    0.923$\pm$0.018&    2.580$\pm$0.096  \\
NGC 6752&        -5.130&   -5.711&   -5.600&   -6.409&   -7.086&    0.693$\pm$0.024&    1.956$\pm$0.079  \\
NGC 6779&        -6.398&   -6.940&   -6.811&   -7.598&   -8.207&    0.671$\pm$0.016&    1.809$\pm$0.079  \\
NGC 6809&        -4.724&   -5.388&   -5.290&   -6.189&   -6.892&    0.763$\pm$0.029&    2.167$\pm$0.127  \\
NGC 6838&        -3.447&   -4.700&   -4.867&   -5.985&   -6.743&    1.086$\pm$0.042&    3.296$\pm$0.165  \\
NGC 6934&        -5.920&   -6.683&   -6.626&   -7.496&   -8.141&    0.820$\pm$0.015&    2.221$\pm$0.073  \\
NGC 6981&        -5.242&   -6.020&   -5.920&   -6.738&   -7.368&    0.878$\pm$0.015&    2.126$\pm$0.076  \\
NGC 7078&        -7.549&   -8.076&   -7.919&   -8.729&   -9.361&    0.685$\pm$0.013&    1.812$\pm$0.051  \\
NGC 7089&        -7.365&   -7.954&   -7.836&   -8.705&   -9.332&    0.709$\pm$0.008&    1.967$\pm$0.040  \\
NGC 7099&        -5.405&   -5.958&   -5.808&   -6.548&   -7.161&    0.702$\pm$0.020&    1.756$\pm$0.086  \\
\enddata
            \label{56GCs}

\end{deluxetable*}


\acknowledgments
We thank the anonymous referee for  a constructive report that has improved the quality of the manuscript.
This work has received funding from the European Research Council (ERC) under the European Union's Horizon 2020 research innovation programme (Grant Agreement ERC-StG 2016, No 716082 'GALFOR', PI: Milone, http://progetti.dfa.unipd.it/GALFOR).
APM and MT acknowledge support from MIUR through the FARE project R164RM93XW SEMPLICE (PI: Milone). APM, MT and ED  have been supported by MIUR under PRIN program 2017Z2HSMF (PI: Bedin). YWL thanks the National Research Foundation of Korea for supporting this research through the grant 2017R1A2B3002919.

\bibliography{ms}{}
\bibliographystyle{aasjournal}

\end{document}